\documentstyle[12pt]{article}
\catcode`\@=11

\@addtoreset{equation}{section}

\def\@normalsize{\@setsize\normalsize{15pt}\xiipt\@xiipt
\abovedisplayskip 14pt plus3pt minus3pt%
\belowdisplayskip \abovedisplayskip
\abovedisplayshortskip  \z@ plus3pt%
\belowdisplayshortskip  7pt plus3.5pt minus0pt}
\def\small{\@setsize\small{13.6pt}\xipt\@xipt
\abovedisplayskip 13pt plus3pt minus3pt%
\belowdisplayskip \abovedisplayskip
\abovedisplayshortskip  \z@ plus3pt%
\belowdisplayshortskip  7pt plus3.5pt minus0pt
\def\@listi{\parsep 4.5pt plus 2pt minus 1pt
            \itemsep \parsep
            \topsep 9pt plus 3pt minus 3pt}}

\def\underline#1{\relax\ifmmode\@@underline#1\else
        $\@@underline{\hbox{#1}}$\relax\fi}
\@twosidetrue
\relax

\catcode`@=12

\catcode`\@=11

\def\section{\@startsection{section}{1}{\z@}{3.5ex plus 1ex minus
   .2ex}{2.3ex plus .2ex}{\large\bf}}

\def\ps@headings{\def\@oddfoot{}\def\@evenfoot{}
\def\@oddhead{\hbox{}\hfill
        \makebox[.5\textwidth]{\raggedright\ignorespaces --\thepage{}--
        \hfill }}
\def\@evenhead{\@oddhead}
\def\subsectionmark##1{\markboth{##1}{}}
}

\ps@headings

\catcode`\@=12

\relax

\def\figcap{\section*{Figure Captions\markboth
        {FIGURECAPTIONS}{FIGURECAPTIONS}}\list
        {Fig. \arabic{enumi}:\hfill}{\settowidth\labelwidth{Fig. 999:}
        \leftmargin\labelwidth
        \advance\leftmargin\labelsep\usecounter{enumi}}}
 \relax
\def\tablecap{\section*{Table Captions\markboth
        {TABLECAPTIONS}{TABLECAPTIONS}}\list
        {Table \arabic{enumi}:\hfill}{\settowidth\labelwidth{Table 999:}
        \leftmargin\labelwidth
        \advance\leftmargin\labelsep\usecounter{enumi}}}
 \relax
\def\reflist{\section*{References\markboth
        {REFLIST}{REFLIST}}\list
        {[\arabic{enumi}]\hfill}{\settowidth\labelwidth{[999]}
        \leftmargin\labelwidth
        \advance\leftmargin\labelsep\usecounter{enumi}}}
 \relax

\catcode`\@=11

\def\ps@headings{\def\@oddfoot{}\def\@evenfoot{}
\def\@oddhead{\hbox{}\hfill
        \makebox[.5\textwidth]{\raggedright\ignorespaces --\thepage{}--
        \hfill }}
\def\@evenhead{\@oddhead}
\def\subsectionmark##1{\markboth{##1}{}}
}

\ps@headings

\relax

\def\firstpage#1#2#3#4#5#6{
\begin{document}
\begin{titlepage}
\nopagebreak
\title{\begin{flushright}
        \vspace*{-1.8in}
        {\normalsize CERN-TH/95-43} -
        {\normalsize UCLA 95/TEP/7}\\[-3mm]
        {\normalsize NUB--#1 --
        #2\\[-9mm]CPTH--RR352.0395}\\[-9mm]
        {\normalsize hep-th/9504034}\\[4mm]
\end{flushright}
\vfill
{\large \bf #3}}
\author{\large #4 \\[1cm] #5}
\maketitle
\vskip -7mm
\nopagebreak
\begin{abstract}
{\noindent #6}
\end{abstract}
\vfill
\begin{flushleft}
\rule{16.1cm}{0.2mm}\\[-3mm]
$^{\star}${\small Research supported in part by\vspace{-4mm}
the National Science Foundation under grant
PHY--93--06906, in part by the EEC contracts \vspace{-4mm}
SC1--CT92--0792, CHRX-CT93-0340 and SC1*CI92--0789, in part by the
U.S. DOE grants DE-AC0381-ER50050 and DOE-AT03-88ER40384, Task E,\vspace{-4mm}
and in part by CNRS--NSF
grant INT--92--16146.}\\[-3mm]
$^{\dagger}${\small Laboratoire Propre du CNRS UPR A.0014.}\\
March 1995
\end{flushleft}
\thispagestyle{empty}
\end{titlepage}}
\newcommand{\dal}{\raisebox{0.085cm}
{\fbox{\rule{0cm}{0.07cm}\,}}}
\newcommand{\dt}{\partial_{\langle T\rangle}}
\newcommand{\dtbar}{\partial_{\langle\bar{T}\rangle}}
\newcommand{\al}{\alpha^{\prime}}
\newcommand{\mst}{M_{\scriptscriptstyle \!S}}
\newcommand{\mpl}{M_{\scriptscriptstyle \!P}}
\newcommand{\dv}{\int{\rm d}^4x\sqrt{g}}
\newcommand{\lv}{\left\langle}
\newcommand{\rv}{\right\rangle}
\newcommand{\ph}{\varphi}
\newcommand{\sbar}{\,\bar{\! S}}
\newcommand{\xbar}{\,\bar{\! X}}
\newcommand{\fbar}{\,\bar{\! F}}
\newcommand{\zbar}{\,\bar{\! Z}}
\newcommand{\dbar}{\,\bar{\!\partial}}
\newcommand{\tbar}{\bar{T}}
\newcommand{\ubar}{\bar{U}}
\newcommand{\ybar}{\bar{Y}}
\newcommand{\phb}{\bar{\varphi}}
\newcommand{\cm}{Commun.\ Math.\ Phys.~}
\newcommand{\pr}{Phys.\ Rev.\ D~}
\newcommand{\pl}{Phys.\ Lett.\ B~}
\newcommand{\ibar}{\bar{\imath}}
\newcommand{\jbar}{\bar{\jmath}}
\newcommand{\np}{Nucl.\ Phys.\ B~}
\newcommand{\e}{{\rm e}}
\newcommand{\gsi}{\,\raisebox{-0.13cm}{$\stackrel{\textstyle
>}{\textstyle\sim}$}\,}
\newcommand{\lsi}{\,\raisebox{-0.13cm}{$\stackrel{\textstyle
<}{\textstyle\sim}$}\,}
\date{}
\firstpage{3118}{IC/95/34}
{\large\sc Perturbative Prepotential and Monodromies\\[-4mm]
in $N{=}2$ Heterotic Superstring$^{\star}$}
{I. Antoniadis$^{\,a}$, S. Ferrara$^{b}$, E. Gava$^{c,d}$,
K.S. Narain$^{ d}$ $\,$and$\,$
T.R. Taylor$^{\,e}$}  %\\[-3mm]
{\normalsize\sl
$^a$Centre de Physique Th\'eorique, Ecole Polytechnique,$^\dagger$
F-91128 Palaiseau, France\\[-3mm]
\normalsize\sl
$^b$Theory Division, CERN, 1211 Geneva 23, Switzerland\\[-3mm]
\normalsize\sl
$^c$Instituto Nazionale di Fisica Nucleare, sez.\ di Trieste,
Italy\\[-3mm]
\normalsize\sl $^d$International Centre for Theoretical Physics,
I-34100 Trieste, Italy\\[-3mm]
\normalsize\sl $^e$Department of Physics, Northeastern
University, Boston, MA 02115, U.S.A.}
{We discuss the prepotential describing the effective field theory of $N{=}2$
heterotic superstring models. At the one loop-level the prepotential develops
logarithmic singularities due to the appearance of charged massless states at
particular surfaces in the moduli space of vector multiplets. These
singularities modify the classical duality symmetry group which now
becomes a representation of the fundamental group of the moduli space
minus the singular surfaces. For the simplest
two-moduli case, this fundamental group turns out to be a certain
braid group and we determine the resulting
full duality transformations of the prepotential,
which are exact in perturbation theory.}
\section{Introduction}

A $N{=}2\,$ supersymmetric gauge theory \cite{n2} is completely
defined by its prepotential --
an analytic function of vector superfields. This analytic structure
is very restrictive and can be used to obtain interesting information
about perturbative as well as non-perturbative behaviour of the theory
\cite{dv}.
Recently, Seiberg and Witten \cite{sw1} constructed a complete
solution of the $SU(2)$ model, and their analysis has been
extended to larger gauge groups in refs.\cite{yank}.
The central point of these studies
is the prepotential describing the massless moduli fields whose
vacuum expectation values break the gauge group down to an abelian subgroup.
It is a very interesting question whether some similar methods could
be employed to analyse the moduli space of superstring theories.

$N{=}2$ supersymmetric, (4,4) \cite{bd} orbifold compactifications
of heterotic superstring theory provide some simplest examples
of string moduli spaces analogous to the globally supersymmetric
spaces considered in refs.\cite{sw1,yank}. A special feature of
these models is the existence of $U(1)\otimes U(1)$ gauge group associated
with an untwisted orbifold plane. Such a plane is parametrized
by two complex moduli, $T$ and $U$, of (1,1) and (1,2) type, respectively. The
tree-level duality group which leaves the mass spectrum and interactions
invariant is $O(2,2;Z)$ \cite{du}, which is isomorphic to the product
of $SL(2,Z)_T$ and  $SL(2,Z)_U$ together with the $Z_2$ exchange of
$T$ and $U$.
The $U(1)\otimes U(1)$ gauge group becomes enhanced to
$SU(2)\otimes U(1)$ along the $T=U$ line, and further enhanced to $SO(4)$
or to $SU(3)$ at $T=U=i$ and $T=U=\rho({=}e^{2\pi i/3})$, respectively
\cite{dijk}.
In this work, we first analyse the perturbative dependence of the prepotential
on this type of moduli, and determine its monodromy properties. Because of the
$N{=}2$ non-renormalization theorems this amounts to computing the one-loop
contributions to the prepotential, as all higher loop corrections vanish. At
the one-loop level the prepotential develops logarithmic singularity due to the
appearance of the additional massless states at the enhanced symmetry
subspaces. As a result, we show that the duality group is modified to a
representation of the fundamental group of the 4-dimensional
space obtained by taking the product of the fundamental domains
of the $T$ and $U$ moduli and removing the diagonal locus. One of
the consequences of this modification is that at the quantum
level the $SL(2,Z)_T$ and $SL(2,Z)_U$ duality transformations do
not commute and also that the $T$, $U$ exchange becomes an element
of infinite order.
The monodromies associated with moving a point around the singular
locus generate a normal abelian subgroup of the full monodromy
group depending on 9 integer
parameters. In addition, there is the usual dilaton shift which commutes with
the above duality group.

In $N{=}2$ heterotic superstrings in four dimensions, the $T,U$ moduli together
with the dilaton-axion $S$ modulus belong to vector multiplets, so their
effective field theory is described by a $N{=}2$ supergravity theory
\cite{n2sg}
coupled to these three vector multiplets. At a generic point of the moduli
space and in the absence of charged massless matter (hypermultiplet) states,
the effective field theory which is obtained by integrating out all massive
string states is local. Its underlying geometric structure is
``special geometry" \cite{spec},
the same structure that appears in the discussion of the
moduli sector of superstrings compactified on Calabi-Yau threefolds. The
symplectic structure based on $Sp(2r)$ for rigid Yang-Mills theories with gauge
group $G$ broken to $U(1)^r$ ($r$ being the rank of $G$)
is here extended to
$Sp(2r+4)$, due to the presence of the additional $S$-vector multiplet and the
graviphoton. For a generic (4,4) compactification of the heterotic superstring
on $T_2\times K_3$, we expect 17 moduli ($r=17$) and a symplectic
structure $Sp(38;Z)$. For a general (4,0) compactification one can also obtain
other values of $r$ up to a maximum of 22. The classical moduli space of vector
multiplets in these theories is
$$\left. {SU(1,1)\over U(1)}\right|_{\makebox{dilaton}}\left.
\times ~{O(2,r)\over O(2)\times O(r)}\right/\Gamma$$
where $\Gamma =O(2,r;Z)$. At a generic point of this moduli space the gauge
group is $U(1)^{r+2}$ and there are no massless charged hypermultiplets. As in
the $O(2,2)$ case there are again complex co-dimension 1 surfaces where
either one
of the $U(1)$'s is enhanced to $SU(2)$ and/or some charged matter
hypermultiplets appear. The one-loop prepotential develops logarithmic
singularities near these surfaces. We study the modifications of the duality
group due to these singularities.

This paper is organized as follows. In section 2, we derive the perturbative
prepotential in $N{=}2$ orbifold compactifications of the heterotic superstring
and study its dependence on the $T,U$ moduli associated with
the untwisted plane.
In section 3, we determine the quantum monodromies of the one-loop
prepotential.
These monodromies are further exploited in section 4, by introducing the usual
$N{=}2$ supergravity basis for the fields where
all transformations act linearly.
We thus find that the duality group $O(2,2;Z)$ is extended to a bigger group
which is contained in $Sp(8,Z)$ symplectic transformations and depends on 15
integer parameters. In section 5, we generalize these results to the full
vector moduli space ($r=17$) for arbitrary $N{=}2$ (4,4) compactifications.
In section 6, we discuss generalizations to (4,0) compactifications. We also
give an explicit orbifold example of two moduli $T,U$ of the untwisted 2-torus
$T^2$, where the orbifold group acts as shifts on the $T^2$. In this case one
encounters singularities associated with the appearance of charged massless
hypermultiplets, as well.
Finally, section 7 contains concluding remarks.

\section{String computation of the one-loop prepotential}

The simplest way to determine the one-loop correction
to the prepotential is to reconstruct it from the K\"ahler
metric of moduli fields. Indeed, the K\"ahler potential of
a $N{=}2$ locally supersymmetric theory can be written as
\begin{equation}
K=-\ln (iY)\: ,~~~~~~~~~~Y=2F-2\bar{F}-{\textstyle\sum_Z}(Z-\bar{Z})
(F_Z+\bar{F}_Z)\: ,      \label{Y}
\end{equation}
where $F$ is the analytic prepotential, $F_Z\equiv\partial_Z F$,
and the summation extends over all chiral ($N{=}2$ vector) superfields
$Z$ \cite{n2sg}. The part of the prepotential that depends on the moduli
of the untwisted plane can be written as
\begin{equation}
F~=~STU+f(T,U)\, ,\label{f}
\end{equation}
where the first term proportional to the dilaton, is the tree-level
contribution, and the one-loop correction is contained in a
dilaton-independent function $f(T,U)$. In our conventions $S$ is defined such
that $\langle S\rangle
={\theta\over\pi}+i{8\pi\over g^2}$ where $g$ is the string
coupling constant and $\theta$ the usual $\theta$-angle.
Thus the one loop moduli metric is
\begin{equation}
K^{(1)}_{Z\bar{Z}} ~=~ \frac{2i}{S-\sbar}G_{Z\bar{Z}}^{(1)}
\end{equation}
with
\begin{equation}
G_{T\tbar}^{(1)}~=~\frac{i}{2(T-\tbar)^2}(\partial_T-\frac{2}{T-\tbar})
(\partial_U-\frac{2}{U-\ubar})f ~+~\makebox{c.c.}  \label{G2}
\end{equation}
and similar expressions for other components. Our first goal is to extract
the function $f(T,U)$ from the moduli metric
obtained in ref.\cite{yuka} by means of a direct superstring computation.

In ref.\cite{yuka}, the $G_{T\tbar}^{(1)}$ component of the metric has
been written as
\begin{equation}
G_{T\tbar}^{(1)}= {\cal I}\,   G_{T\tbar}^{(0)}\, ,     \label{GI}
\end{equation}
where $G_{T\tbar}^{(0)}=- (T-\tbar)^{-2}$ is the tree-level metric,\footnote{
A $Z\rightarrow iZ$ rescaling on the chiral fields of ref.\cite{yuka}
is necessary to recover the chiral fields as defined here.}
and the world-sheet integral
\begin{equation}
{\cal I}  = \int\frac{d^2\tau}{\tau_{2}^{\; 2}}\bar{F}(\bar{\tau})\,
\partial_{\bar{\tau}}(\tau_2 \sum_{p_L,p_R}e^{\pi i\tau\, |p_{L}|^2}\,
e^{-\pi i \bar{\tau\,} |p_{R}|^2})                     \label{I1}
\end{equation}
extends over the fundamental domain of
the modular parameter $\tau\equiv\tau_1+i\tau_2$.
In eq.(\ref{I1}), $\bar{F}(\bar{\tau})=\overline{F(\tau)}$,
where $F(\tau)$ is a moduli-independent meromorphic form of weight
$-$2 with a simple pole at infinity due to the tachyon of the bosonic
sector. This in fact fixes $F$ completely up to a multiplicative
constant:
\begin{equation}
F(\tau)=-{1\over\pi}\,{j(\tau)[j(\tau)-j(i)]\over j_{\tau}(\tau)}
\label{Ftau}
\end{equation}
where $j$ is the meromorphic function with a simple pole with residue 1 at
infinity and a third order zero at $\tau=\rho$.
The summation inside the integral extends over the left- and right-moving
momenta in the untwisted orbifold plane. These momenta are parametrized as
\begin{eqnarray}
p_L &=& \frac{1}{\sqrt{2\,{\makebox Im}T\,{\makebox Im}U}}\,
(m_1+m_2\ubar+n_1\tbar+n_2\tbar\ubar)                        \label{pL}\\
p_R &=& \frac{1}{\sqrt{2\,{\makebox Im}T\,{\makebox Im}U}}\,
(m_1+m_2\ubar+n_1T+n_2T\ubar)                                 \label{pR}
\end{eqnarray}
with integer $m_1$, $m_2$, $n_1$ and $n_2$.

In ref.\cite{yuka} it has been shown that the integral $\cal I$ satisfies
the differential equation
\begin{equation}
[\partial_T\partial_{\tbar}+ \frac{2}{(T-\tbar)^2}]{\cal I}=
-{4\over (T-\tbar)^2} \int d^2\tau \bar{F}(\bar{\tau})\,\partial_{\tau}
(\partial_{\bar{\tau}}^2 +{i\over\tau_2}\partial_{\bar{\tau}})(\tau_2
\sum_{p_L,p_R}e^{\pi i\tau\, |p_{L}|^2}\, e^{-\pi i \bar{\tau\,} |p_{R}|^2})
\, .
\label{eq}
\end{equation}
The r.h.s.\ being a total derivative with respect to $\tau$ vanishes away from
the enhanced symmetric points $T=U$ (modulo $SL(2,Z)$). However, as it has been
pointed out by Kaplunovsky \cite{kapl},
the surface term gives rise to a $\delta$-function
due to singularities associated with the additional massless particles at
$T=U$.
They correspond to lattice momenta (\ref{pL}), (\ref{pR}) with
$m_1=n_2=0$ and $m_2=-n_1=\pm 1$, so that $p_L=0$ and $p_R=\pm i\sqrt{2}$.
These are the two additional gauge multiplets which enhace the gauge symmetry
to $SU(2)\times U(1)$.
Expanding $p_L$, $p_R$ around $T=U$ for these states, it is easy to show that
the surface term becomes proportional to:
$$
\lim_{\tau_2\rightarrow\infty} \tau_2 e^{-{\pi\tau_2 |T-U|^2\over 2{\makebox
Im}T {\makebox Im}U}}\ \sim\ \delta^{(2)} (T-U) \ .
$$
Note that there are two special points on the $T=U$
plane (modulo $SL(2,Z)$) where the gauge symmetry is further enhanced: $T=U=i$
giving rise to $SO(4)$ and $T=U=\rho$ to $SU(3)$, $\rho$ being the cubic root
of unity. We will comment on these special points later.
To solve eq.(\ref{eq}) we will stay away from the singular region and we will
take into account the singularity structure by suitable boundary conditions.
We therefore have the following equations:
\begin{equation}
[\partial_T\partial_{\tbar}+ \frac{2}{(T-\tbar)^2}]{\cal I}=
[\partial_U\partial_{\ubar}+ \frac{2}{(U-\ubar)^2}]{\cal I}=0 \, .
\label{equ}
\end{equation}

The general solution of eqs.(\ref{equ}) is
\begin{equation}
{\cal I}={1\over 2i}(\partial_T-\frac{2}{T-\tbar})
[(\partial_U-\frac{2}{U-\ubar})f(T,U)
+(\partial_{\ubar} + \frac{2}{U-\ubar})\tilde{f}(T,\ubar)] + {\rm c.c.}
\, ,
\label{sol}
\end{equation}
where $f$ and $\tilde{f}$ depend only on the indicated variables. The above
equation is not in the form (\ref{G2}) dictated by $N{=}2$ supersymmetry due to
the presence of $\tilde{f}$ but we will now show that the latter vanishes.
Taking
appropriate derivatives of eq.(\ref{sol}) one finds the following identity:
\begin{equation}
D_{\ubar} \partial_{\ubar} D_T\partial_T {\cal I} = {1\over 2i}\partial_T^3
\partial_{\ubar}^3 \tilde{f}\ ,
\label{ftil}
\end{equation}
where the covariant derivative $D_T = \partial_T + {2\over T-\tbar}$. Now we
can evaluate the l.h.s.\ of the above equation by using the explicit string
expression (\ref{I1}) for $\cal{I}$ with the forms (\ref{pL}) and (\ref{pR})
for the lattice momenta. The result is:
\begin{equation}
\partial_T^3 \partial_{\ubar}^3 \tilde{f}=
-{16\pi^2\over (T-\tbar)^2 (U-\ubar)^2}
\int\frac{d^2\tau}{\tau_{2}^{\; 2}}\bar{F}(\bar{\tau})\,
\partial_{\bar{\tau}}(\tau_2^2 \partial_{\tau} (\tau_2^2 \partial_{\tau}
( \tau_2 \sum_{p_L,p_R} {\bar{p}}_R^4 e^{\pi i\tau\, |p_{L}|^2}\,
e^{-\pi i \bar{\tau\,} |p_{R}|^2})))
\label{ftil1}
\end{equation}
One can show that the r.h.s.\ is a total derivative in $\tau$ and
vanishes away from the enhanced symmetric points. As a result, the general
solution for $\tilde{f}$ is a quadratic polynomial in $T$ and $\ubar$.
However such a polynomial can be reabsorbed in the function $f(T,U)$, as can
be seen from the expression (\ref{sol}) for $\cal{I}$. Therefore without loss
of generality we can set $\tilde{f}=0$.
This result is compatible with $N{=}2$ supersymmetry, as seen
from eqs.(\ref{GI}) and (\ref{G2}) and the function $f$ appearing in
(\ref{sol}) can be identified with the one loop correction to the
prepotential (\ref{f}).

Our next task is to determine $f$.
Equation (\ref{sol}) has no simple holomorphic structure, therefore
it is not suitable for exploiting the holomorphy
property of the prepotential.
However, a simpler equation can be obtained by taking appropriate
derivatives as in the case of $\tilde{f}$ above. It can be shown that
\begin{equation}
-i(U-\ubar)^2 D_T\partial_T \partial_{\ubar} {\cal I} = \partial_T^3 f\ .
\label{op}
\end{equation}
A straightforward calculation utilizing eqs.(\ref{pL}) and (\ref{pR}) yields
\begin{equation}
f_{TTT}=8\pi^2 \frac{U-\ubar}{(T-\tbar)^2}
\int\frac{d^2\tau}{\tau_{2}^{\; 2}}\bar{F}(\bar{\tau})\,
\partial_{\bar{\tau}}[{\tau_{2}^{\; 2}}\partial_{\tau}({\tau_{2}^{\; 2}}
\sum_{p_L,p_R}p_L\bar{p}_{R}^3\, e^{\pi i\tau\, |p_{L}|^2}\,
e^{-\pi i \bar{\tau}\, |p_{R}|^2})]\, .
\end{equation}
The r.h.s.\ can be further simplified by integrating by parts. The boundary
term is vanishing away from the enhanced symmetry points and the result is:
\begin{equation}
f_{TTT}=4\pi^2 \frac{U-\ubar}{(T-\tbar)^2}
\int{d^2\tau}\,\bar{F}(\bar{\tau})\,
\sum_{p_L,p_R}p_L\bar{p}_{R}^3\, e^{\pi i\tau\, |p_{L}|^2}\,
e^{-\pi i \bar{\tau}\, |p_{R}|^2}\, .
\label{fTTT}
\end{equation}
The r.h.s.\ of the above equation is indeed an analytic function of
$T$ and $U$, as can be verified by taking derivatives with respect to $\tbar$
or $\ubar$. The resulting expressions are total derivatives in $\tau$ and
vanish upon integration.

We now employ the $SL(2,Z)_T\otimes SL(2,Z)_U$ spacetime duality symmetry
in order to further determine the r.h.s.\ of (\ref{fTTT}).
Under $SL(2,Z)_T$ transformations,
\begin{equation}
T\rightarrow {aT+b \over cT+d}\ ,
\label{sl2z}
\end{equation}
the lattice momenta (\ref{pL}), (\ref{pR}) transform as $(p_L, {\bar p}_R)
\rightarrow ((cT+d)/(c\tbar +d))^{1/2} (p_L, {\bar p}_R)$ modulo relabeling of
the integers $m_i,n_i$. Similarly under
$SL(2,Z)_U$ transformations, they transform as $(p_L, p_R)
\rightarrow ((cU+d)/(c\ubar +d))^{1/2} (p_L, p_R)$. Using these properties one
can verify that the r.h.s.\ of eq.(\ref{fTTT}) behaves like a
meromorphic modular function of weight 4 in $T$ and $-$2 in $U$. Furthermore,
the only sigularity in the $T,U$ plane (including infinities) is a simple pole
at $T=U$ (modulo $SL(2,Z)_U$). Indeed, by expanding $p_L$ and $p_R$ around
$T=U$
for the additional massless states, one finds that the r.h.s.\ behaves like
$\int d\tau_2 (\tbar-\ubar) e^{-{\pi\tau_2 |T-U|^2\over 2{\makebox Im}T
{\makebox Im}U}}\ \sim\ {1/(T-U)}$. Following the standard theorems of modular
forms, we find
\begin{equation}
f_{TTT}=\frac{j(U)\,[j(U)-j(i)]}{j_U(U)\,[j(U)-j(T)]}\,h(T)\, , \label{umod}
\end{equation}
where $j$ is defined below eq.(\ref{Ftau}) and $h(T)$ is a meromorphic modular
function of weight 4, with at most a first order pole at infinity. Inspection
of the integral (\ref{fTTT}) shows that $f_{TTT}\rightarrow 0$ as $T\rightarrow
i\infty$ which implies that $h(T)$ must be holomorphic everywhere. This
therefore fixes $f_{TTT}$ uniquely to:\footnote{This
result was also known to V.\ Kaplunovsky \cite{kapl}, as
recently reported by B. de Wit, V. Kaplunovsky, J. Louis and
D. L\"ust in preprint hep-th/9504006.}
\begin{equation}
f_{TTT}=-{2i\over\pi}\frac{j_T(T)}{j(T)-j(U)}
\left\{\frac{j(U)}{j(T)}\right\}
\left\{\frac{j_T(T)}{j_U(U)}\right\}
\left\{\frac{j(U)-j(i)}{j(T)-j(i)}\right\}\equiv 2 W(T,U).      \label{f3T}
\end{equation}

The function $f_{UUU}$ is obtained from eq.(\ref{f3T})  by replacing
$T\leftrightarrow U$. A tedious calculation shows that the result is consistent
with the integrability condition
\begin{equation}
\partial_{U}^3f_{TTT}=\partial_{T}^3f_{UUU}\, , \label{int}
\end{equation}
which is necessary for the existence of the prepotential $f(T,U)$.

In order to find a solution $f$ for the above differential equations,
it is convenient to introduce the following closed meromorphic one-form
$\omega$:
\begin{equation}
\omega(T,U;T',U')=dT'Q(U,U')(T-T')^2W(T',U')+dU'Q(T,T')(U-U')^2W(U',T'),
\label{omega}
\end{equation}
where $Q(x,x')$ is the second order differential operator defined
as:
\begin{equation}
Q(x,x')=\frac{1}{2}(x-x')^2{\partial}^2_{x'}+
(x-x'){\partial}_{x'}+1.
\label{diff}
\end{equation}
Using the property ${\partial}_{x'}Q(x,x')=
\frac{1}{2}(x-x')^2{\partial}^3_{x'}$ and the integrability condition
(\ref{int}), one can indeed prove that $\omega$ is closed, namely:
$d'\omega=0$, where $d'\equiv dU'\partial_{U'}+dT'\partial_{T'}$.
For non-singular $(T,U)$, one can show that
the following  line
integral of $\omega$ satisfies the differential equations for $f(T,U)$,
therefore defining the latter up to a quadratic polynomial in $T$ and $U$:
\begin{equation}
f(T,U)=\int^{(T,U)}_{(T^0,U^0)}\omega(T,U;T',U'),
\label{intf}
\end{equation}
where $(T^0,U^0)$ is an arbitrary base point (outside the singular
locus of $\omega$), different choices of the base point modifying $f(T,U)$
by a quadratic polynomial, as is evident from the fact that $\omega$ is
quadratic in $T,U$.
The path of integration in (\ref{intf}) is chosen such that
it does not cross any singularity.
Note that the complement of the singular
locus is connected and therefore such a path always exists,
however this complement is not simply connected, and as a result the above
line integral depends on the homology class of the integration path. Different
choices of homology classes of paths will alter $f$ by quadratic polynomials
in $T,U$.
This ambiguity is related to the non-trivial quantum monodromies which will be
discussed in the next section.

The other important point concerns the transformation properties
of $f(T,U)$ under the action of $PSL(2,Z)$ on $T$ and $U$. From the equation
defining $\omega$, it follows that under $T\rightarrow
T_g\equiv\frac{aT+b}{cT+d}$ we have:
\begin{equation}
\omega(T_g,U;T'_g,U')=(cT+d)^{-2}\omega(T,U;T',U').
\label{omegat}
\end{equation}
Using this property in (\ref{intf}) one can derive the following
equation:
\begin{equation}
f(T_g,U)=(cT+d)^{-2}[f(T,U)+\int_{(T^0_{g^{-1}},U^0)}^{(T^0,U^0)}
\omega(T,U;T',U')].
\label{intft}
\end{equation}
The homology class of path defining
the second term of the r.h.s.\ of this equation is determined by those
defining $f(T,U)$ and $f(T_g,U)$. We will be more precise on this point
in the next section, however we note here
that equation (\ref{intft}) implies that $f$ transforms with weight $-2$
in $T$ up to
a quadratic polynomial in $T,U$ coming from the second term in the r.h.s.\
of (\ref{intft}). The same transformation properties hold for
the $U$ variable.
Similarly under $T,U$ exchange one can show that:
\begin{equation}
f(U,T)=f(T,U)+\int_{(U^0,T^0)}^{(T^0,U^0)}\omega(T,U;T',U'),
\label{intfex}
\end{equation}
implying again that $f$ picks an additive quadratic polynomial.

When $U$ is one of the fixed points of the modular group $SL(2,Z)_U$ (e.g.\
the order 2 fixed point $U=i$ or the order 3 fixed point $U=\rho$), $f_{TTT}$
vanishes. Let us consider the behaviour of $f_{TTT}$ at generic $U$ away
from these fixed points.
As mentioned above, eq.(\ref{f3T}) is
singular as $T$ approaches $U_g =
\frac{aU+b}{cU+d}$ where $g$ is an $SL(2,Z)$ element:
\begin{equation}
f_{TTT} \rightarrow -{2i\over\pi}\frac{1}{T-U_g} (cU+d)^2\ .\label{lim}
\end{equation}
Note that if $U_g$ is one of the fixed points then one must sum over the
residues around the poles $1/(T-U_{gg'})$ where $g'$ is an element of
the little group of $U_g$. It is easy to verify that the resulting sum vanishes
consistent with the fact that $f_{TTT}$ is zero at these points.
Upon integration, the limit (\ref{lim}) becomes
\begin{equation}
f(T,U) \rightarrow -{i\over\pi}[(cU+d)T-(aU+b)]^2\ln(T-U_g)\, ,  \label{limf}
\end{equation}
giving rise to a branch cut starting
at $T=U_g$. When $U_g$ is not one of
the fixed points, it follows from eq.(\ref{G2}) that
\begin{equation}
G_{T\tbar}^{(1)}\rightarrow {1\over\pi} \ln|T-U_g|^2\, G_{T\tbar}^{(0)}\ .
\label{limg}\end{equation}
When $U_g$ is one of the fixed points then the summation over the little group
of $U_g$ introduces a multiplicative factor 2 or 3 for the fixed points of
order 2 or 3, corresponding to the enhanced symmetries $SO(4)$ or
$SU(3)$ respectively.

The singular behaviour (\ref{limg}) of the modulus
(and its $N{=}2$ superpartners) wave function renormalization factor
can be understood within the framework of effective
field theory. It is due to infrared divergences which arise
in the presence of massless particles carrying non-zero charges
with respect to the $U(1)$ gauge group associated with the $N{=}2$ vector
multiplet of $T$. The field-theoretical result is
\begin{equation}
G_{T\tbar}^{(1)}\rightarrow \frac{1}{2\pi}\sum_a e_{a}^{2}\ln m_{a}^{2}\,
G_{T\tbar}^{(0)}\ ,
\label{limft}\end{equation}
where $e_a$ and $m_a \propto|T-U_g|$ are the charges and masses, respectively,
of $N{=}2$ vector multiplets
that become massless in the $T\rightarrow U_g$ limit.
These multiplets do indeed carry non-zero charges, and it is not difficult
to show that eq.(\ref{limft}) agrees with eq.(\ref{limg}). The multiplicative
factors of 2 and 3 at the fixed points of order 2 and 3 respectively arise
due to the presence of additional charged massless states corresponding
to the gauge groups $SO(4)$ and $SU(3)$. Indeed, the ratio 1:2:3 corresponds
the the ratio of 1/2 of the $SU(2)$ $\beta$-function to the $\beta$-functions
of
$SO(4)$ and $SU(3)$. The factor 1/2 is due to the fact that the field which
has well-defined quantum numbers under $SU(2)$ is not $T$ itself but
the combination $(T-U)$.

\section{Monodromies of the one-loop prepotential}

Now we turn to the question of the monodromy group that acts on $f$.
At the classical level there is
the usual action of the modular group acting on $T$ and $U$ upper half planes,
namely $PSL(2,Z)_T\otimes PSL(2,Z)_U$.   The $PSL(2,Z)_T$ subgroup of the
$PSL(2,Z)_T\otimes PSL(2,Z)_U$ modular symmetry group
is generated by the transformations
\begin{equation}
g_1: ~T\rightarrow -1/T ~~~~~~~~~~~~~g_2: ~T\rightarrow -1/(T+1)\ .  \label{A}
\end{equation}
The $PSL(2,Z)_U$ subgroup is generated by
\begin{equation}
g'_1: ~U\rightarrow -1/U~~~~~~~~~~~~~g'_2: ~U\rightarrow -1/(U+1)\ .  \label{B}
\end{equation}
These generators obey the $SL(2,Z)$ relations
\begin{equation}
(g_1)^2 = (g'_1)^2 = (g_2)^3 = (g'_2)^3 = 1\ ,
\label{grel}
\end{equation}
and the relations implied by the fact that the two $PSL(2,Z)$'s commute.
There is also an exchange symmetry generator, namely:
\begin{equation}
\sigma: ~T\leftrightarrow U,
\label{C}
\end{equation}
which satisfies $\sigma^2=1$. Moreover $\sigma$ relates the two
$PSL(2,Z)$'s via $g'_1=\sigma g_1\sigma$ and $g'_2=\sigma g_2\sigma$.
We expect that these relations do not hold in the quantum case, due
to the singularities of the prepotential. For instance, since
$\sigma^2$ corresponds to moving a point around $T=U$
singularity, it will not be equal to the identity.
In order to understand the monodromy properties in the quantum
case we have to find the new relations among the generators.
To do that it is convenient to think of the above relations
as relations among the generators of the fundamental group
of the underlying moduli space. The classical monodromy group is
then obtained by imposing the relation $\sigma^2=1$, while
in the quantum case this relation is modified by the presence
of a logarithmic branch cut.

At the classical level the underlying space is
the product of two $PSL(2,Z)$ fundamental
domains with an identification given by $\sigma$. Topologically each of
these two fundamental domains can be thought of as a two-sphere
$S$ ($S'$) with $3$ distinguished points $x_1$ ($x'_1$),
$x_2$ ($x'_2$) and $x_3$ ($x'_3$),
which can be taken to be the images
of $i$, $\rho$ and $\infty$ by the $j$-function.
Associated with these three points
we have generators $g_i$ ($g'_i$) of the fundamental group of orders $2$,
$3$ and $\infty$ respectively, subject to the conditions
$g_3g_2g_1=1$ and $g'_3g'_2g'_1=1$. The total space
is then the product of the two spheres $S$ and $S'$ minus $\{x_i\}\times S'$
and $S\times\{x'_i\}$, $i=1,2,3$, and the fundamental
group of the resulting $4$-dimensional space is the product of the
fundamental groups of the two punctured spheres. Including $\sigma$,
we have the additional relations $g'_i=\sigma g_i \sigma$ and $\sigma^2=1$.

In the quantum case however, since
we have singularities at $T=U$, we must remove the diagonal in the
product of the two punctured spheres and this modifies the structure of the
fundamental group. In general, when one takes a product of two (or more)
identical Riemann surfaces and removes the diagonal, the fundamental group
of the resulting space is called braid group and has been studied
extensively \cite{Scott}. One can adapt the results
of ref.\cite{Scott} to the present case, and obtain the following
relations:
\begin{eqnarray}
&&g_3 g_2 g_1=\sigma^2,~~~~~~(g_1)^2=(g_2)^3=1\nonumber\\
&&g'_i=\sigma^{-1}g_i\sigma\nonumber\\
&&g_1\sigma^{-1}g_2\sigma=\sigma^{-1}g_2\sigma g_1\nonumber\\
&&\sigma g_i\sigma^{-1}g_i=g_i\sigma^{-1}g_i\sigma.
\label{qrelations}
\end{eqnarray}
The full fundamental group is indeed generated by three elements
$\sigma$, $g_1$, $g_2$ subject to the above relations.
Notice that if one sets $\sigma^2=1$ one gets back the classical relations
for the two commuting $PSL(2,Z)$'s. However, as mentioned earlier, in the
quantum case $\sigma^2\neq 1$ and the two $PSL(2,Z)$'s do not commute
anymore.
In fact, $\sigma^2$ corresponds to moving a point around the singularity
at $T=U$ and therefore transforms the prepotential $f$
non-trivially:
\begin{equation}
Z_1 \equiv \sigma^2 : f(T,U) \rightarrow f(T,U) +2 (T-U)^2
\label{Z1}
\end{equation}
Note that the additive piece above is uniquely fixed by the fact that
it must be at most quadratic in $T$ as well as $U$
and by the behaviour of $f$ near $T=U$ governed by the
logarithmic term in eq.(\ref{limf}).

Actually one can explicitly check the non commutativity
of $T$ and $U$ duality transformations using the integral
representation for $f$ given in (\ref{intf}). For instance
one finds for the commutator $g_1g'_1(g_1)^{-1}(g'_1)^{-1}$:
\begin{equation}
g_1g'_1(g_1)^{-1}(g'_1)^{-1}: f(T,U) \rightarrow f(T,U) +
2(T-U)^2-2(1+TU)^2
\label{comm}
\end{equation}

Notice also that we could redefine $g_3$ in the first equation
of (\ref{qrelations}) by $\tilde{g}_3=\sigma^{-2}g_3$, and then
$\tilde{g}_3g_2g_1=1$, which is the usual $SL(2,Z)$ relation.
We can do the same for $g'_i$, showing that the quantum monodromy
group contains the two $SL(2,Z)$'s as subgroups. However,
as seen from (\ref{qrelations}) the two $SL(2,Z)$'s now do not commute.

Having the generators and relations of the fundamental group,
we will now determine the monodromy transformations of the
prepotential $f$. We can assume the following transformation
properties of $f$ under the generators $g_1$, $g_2$ and $\sigma$:
\begin{eqnarray}
g_1: &~~ T\rightarrow -1/T ~~;~~& f\rightarrow T^{-2}\,( f+P(T,U)\, )\ ,
\nonumber\\
g_2: &~~ T\rightarrow -1/(T+1) ~~;~~& f\rightarrow (T+1)^{-2}\,(
f+R(T,U)\, )\ ,
\nonumber\\
\sigma: &~~ T\leftrightarrow U~~;~~&f\rightarrow f+K(T,U)\ ,
\label{transform}
\end{eqnarray}
As explained in the previous section the functions $P$, $R$ and
$K$ are polynomials quadratic in $T$ and $U$. Note that this property is
consistent with the requirement that the quantity ${\cal I}$ which gives the
physical metric (\ref{GI}) remains invariant under all three transformations.
In fact, using eq.(\ref{sol}), one finds that these functions must satisfy
$$
{\rm Im}\{(\partial_T-{2\over T-\tbar})(\partial_U-{2\over U-\ubar})Q\}=0 ~~;~~
Q\equiv P,R,K
$$
It is then straightforward
to show that the most general solution to this equation is a
general quadratic polynomial in both $T,U$ with real coefficients.

The functions $P,R,K$ must be compatible with the relations
(\ref{qrelations}) and also with (\ref{Z1}). The latter implies that:
\begin{equation}
K(T,U) + K(U,T) = 2 (T-U)^2.
\end{equation}
The general solution for $K(T,U)$ then is:
\begin{equation}
K(T,U) = (T-U)^2 + (T-U)(xUT + y(T+U) +z),
\label{K1}
\end{equation}
where $x$, $y$ and $z$ are complex numbers.
The relation $(g_1)^2=1$ implies that $P$ must be of the
form $\alpha(T^2-1)+\beta T$ where $\alpha$ and $\beta$
are quadratic polynomials in $U$. Similarly from the relation $(g_2)^3=1$
one finds that $R=AT^2+2(A+C)T+C$, with $A$ and $C$ quadratic in
$U$. Using the freedom to add to $f$ a
quadratic polynomial in $T$ and $U$ (involving 9 parameters)
we can set for example 9 parameters entering in $\alpha$, $\beta$ and $A+C$
to zero. Using the last two relations
of (\ref{qrelations}), we can then show that all the remaining parameters
get fixed, resulting into the following expressions for the 3 polynomials:
\begin{eqnarray}
P&=&0\nonumber\\
R&=&2(T^2-1)\label{poly}\\
K&=&(T-U)^2+(T-U)(-2UT+T+U+2).\nonumber
\end{eqnarray}
Notice that the coefficients of the polynomials are real,
and as a result one can check, using (\ref{G2}), that the K\"ahler
metric transforms covariantly.

The full monodromy group $G$ contains a normal abelian subgroup $H$,
which is generated by elements $Z_g$ obtained by conjugating $Z_1$
by an element $g$ which can be any word in the
$g_i$'s, $g'_i$'s and their inverses.
More explicitly, if
$g$ acts on the $T,U$ space as $T\rightarrow T$ and $U\rightarrow \frac
{aU+b}{cU+d}$, then $Z_g$ acts as:
\begin{equation}
Z_g : (T,U)\rightarrow (T,U)~~;~~ f(T,U) \rightarrow f(T,U)
+2((cU+d)T-(aU+b))^2
\label{Zg}
\end{equation}
In other words $Z_g$ corresponds to moving a point around
the singularity $T=U_g$, where the prepotential behaves as shown
in (\ref{limf}). Notice that the fact that $H$ is abelian does not follow from
the general group structure
of (\ref{qrelations}), but from the specific logarithmic
singularity (\ref{limf}), which implies that $H$
acts on $f$ by shifts as in (\ref{Zg}).
A general element of $H$ is obtained by a sequence of
such transformations and shifts $f$ by:
\begin{equation}
f\rightarrow f +2\sum_i N_i ((c_iU+d_i)T-(a_iU+b_i))^2
\equiv f +\sum_{n,m =0}^2 c_{nm} T^n U^m ~~~ N_i \in Z
\label{He}
\end{equation}
with $a_i,b_i,c_i,d_i$ corresponding to some
$SL(2,Z)$ elements for each $i$.
Since the polynomial entering in (\ref{He}) has 9 independent parameters
$c_{nm}$, it follows that $H$ is isomorphic to $Z^9$.
The set of all conjugations of $H$ by elements generated by $g_i$'s and
$g'_i$'s defines a group of (outer) automorphisms of $H$ which is isomorphic to
$PSL(2,Z)\times PSL(2,Z)$, under which
$c_{nm}$ transform as $(3,3)$ representation (in this notation the two
$PSL(2,Z)$'s act on the index $n,m$ respectively). Moreover, the
conjugation by $\sigma$ defines an automorphism which interchanges the indices
$n$ and $m$ in $c_{nm}$. Thus the set of all conjugations
of $H$ is isomorphic to $O(2,2;Z)$, under which the
$c_{nm}$'s transform as a second rank traceless
symmetric tensor. Finally, the quotient group $G/H$ is isomorphic
to $O(2,2;Z)$, therefore $G$ is a group involving
15 integer parameters. On the other hand,
$G$ is not a semidirect product of $O(2,2;Z)$ and
$H$, since $O(2,2;Z)$ is not a subgroup of $G$, as it follows from the
quantum relations (\ref{qrelations}).
Of course for physical on-shell quantities the group $H$ acts
trivially and therefore one recovers the usual action
of $O(2,2;Z)$.

\section{Linear basis for the monodromies and quantization}

So far we have discussed the monodromies of $f$, which turned out to be
consistent with the covariance of the
K\"ahler metric. However, in order for the K\"ahler
potential
to transform by a K\"ahler transformation, the transformations of $f$ must be
supplemented by  suitable transformations of the dilaton field $S$. From the
form of the K\"ahler potential (\ref{Y}) and (\ref{f}) one deduces that $S$
must
transform as:
\begin{eqnarray}
g_1: S&\rightarrow& S+ {f_U\over T}\\
g_2: S&\rightarrow& S+ {f_{U}\over {T+1}}\\
\sigma: S&\rightarrow& S- {1\over 2}K_{TU}
\label{stran}
\end{eqnarray}
One can verify that the above transformations satisfy all
the group constraints discussed earlier. The above equations therefore
define the action of the monodromy group $G$ on $S$. In addition to this,
there is also the usual axionic shift which leaves $T$, $U$, and $f$ invariant,
\begin{equation}
D: S\rightarrow S + \lambda\ ,
\label{sshift}
\end{equation}
where $\lambda$ is a real number. The full perturbative group of monodromies
is the direct product of $G$ with the abelian translation group
(\ref{sshift}).

In order to better understand the group stucture and discuss quantization of
the parameters due to non-perturbative effects, it is convenient to introduce a
field basis where all monodromies act linearly. To this end we use the
formalism of the standard $N{=}2$ supergravity \cite{n2sg}
where the physical scalar fields
$Z^I$ of vector multiplets are expressed as $Z^I=X^I/X^0$, in terms of the
constrained fields $X^I$ and $X^0$. This is a way to include the extra $U(1)$
gauge boson associated with the graviphoton which has no physical scalar
counterpart. In our case we have
\begin{equation}
S={X^s\over X^0} ~~~ T={X^2\over X^0} ~~~ U={X^3\over X^0}
\label{xbasis}
\end{equation}
and the prepotential (\ref{f}) is the following homogeneous polynomial of
degree 2:
\begin{equation}
F={X^sX^2X^3\over X^0} + (X^0)^2f({X^2\over X^0},{X^3\over X^0})
\label{fxbasis}
\end{equation}
The K\"ahler potential $K$ is
\begin{equation}
K=-\log i(\xbar^I F_I-X^I \fbar_I)\ ,
\label{Kxbasis}
\end{equation}
where $F_I$ is the derivative of $F$ with respect to $X^I$ and $I=0,s,2,3$.
This has a generalization in basis where $F_I$ is not the derivative of a
function $F$ \cite{ceres}. Then, the kinetic matrix for vector fields $N_{IJ}$
is a $4\times 4$ symmetric matrix completely determined by $X^I$ and $F_I$
through the formulae (\ref{Kxbasis}) and
$$
F_I=N_{IJ}X^J ~~~,~~~ {\cal D}_I\fbar_J = N_{JL}{\cal D}_I\xbar^L
$$
where ${\cal D}_I=\partial_I+K_I$. For the case in which $F_I=\partial_I F$, it
reduces to the known expression of ref. \cite{n2sg}.

It is clear that symplectic transformations acting on $(X^I,F_I)$ leave the
K\"ahler potential invariant. Since the monodromy group leaves $K$ invariant,
we expect it to be a subgroup of the symplectic group $Sp(8)$. In the following
we will identify this subgroup. A general symplectic transformation is
\begin{equation}
\pmatrix{X^I\cr F_I\cr}\rightarrow \pmatrix{{\bf a}&{\bf b}\cr {\bf c}&{\bf d}
\cr}
\pmatrix{X^I\cr F_I\cr}
\label{symp}
\end{equation}
where ${\bf a}$, ${\bf b}$, ${\bf c}$, ${\bf d}$ are $4\times 4$ matrices and
satisfy the defining relations of the symplectic group, namely
\begin{equation}
{\bf a}^t {\bf c}- {\bf c}^t{\bf a}=0 ~~~,~~~ {\bf b}^t{\bf d}-{\bf d}^t{\bf b}
=0~~~,~~~{\bf a}^t {\bf d}-{\bf c}^t{\bf b}={\bf 1}.
\label{abcd}
\end{equation}
Under this transformation, however, the vector kinetic term Im${\cal
F}^I_{\mu\nu} {\overline N}_{IJ} {{\cal F}^J}^{\mu\nu}$ transforms as:
\begin{equation}
N \rightarrow ({\bf c}+{\bf d}N)({\bf a}+{\bf b}N)^{-1}.
\label{vector}
\end{equation}
If ${\bf b}\ne 0$ then from the above equation it follows that the gauge
coupling gets inverted and therefore in a suitable basis the perturbative
transformations must have ${\bf b}=0$. When ${\bf b}=0$ the symplectic
contraints (\ref{abcd}) imply that ${\bf d}^t={\bf a}^{-1}$ and ${\bf c}=
{{\bf a}^t}^{-1}{\tilde{\bf c}}$ with ${\tilde{\bf c}}$ an arbitrary symmetric
matrix. Furthermore, from eq.(\ref{vector}) we see that the vector kinetic
term changes by ${\tilde{\bf c}}_{IJ}\makebox{Im}{\cal F}^I{\cal F}^J$ which,
being a total derivative, is irrelevant at the perturbative level. However at
the non-perturbative level, due to the presence of monopoles, the matrix
${\tilde{\bf c}}$ must have integer entries.

In the absence of the one-loop correction $f$, one can verify that the
$PSL(2,Z)_T$ transformation $T\rightarrow{aT+b\over cT+d}$ transform $X^I$ and
$F_I$ as:
\begin{equation}
\matrix{X^0&\rightarrow&cX^2+dX^0\qquad\qquad &F_0&\rightarrow&aF_0-bF_2\cr
X^s&\rightarrow&cF_3+dX^s\qquad\qquad &F_s&\rightarrow&aF_s+bX^3\cr
X^2&\rightarrow&aX^2+bX^0\qquad\qquad &F_2&\rightarrow&-cF_0+dF_2\cr
X^3&\rightarrow&cF_s+dX^3\qquad\qquad &F_3&\rightarrow&aF_3+bX^s\cr}
\label{treetr}
\end{equation}
and similarly $PSL(2,Z)_U$ transformation is given by interchanging
$X^2$ with $X^3$ and $F_2$ with $F_3$ in the above equation. Note that these
transformations act linearly and are in fact symplectic. However, in this basis
the matrix ${\bf b}\ne 0$ as $X^I$'s get transformed to $F^I$'s. It is
therefore
convenient to make a symplectic change of the basis into $(X^I, F_I)$ where
$I=0,1,2,3$ with $X^1=F_s$ and $F_1=-X^s$. In the new basis the tree-level
$O(2,2;Z)$ transformations are block diagonal, i.e.\ ${\bf b}={\bf c}=0$ and
${\bf d}={{\bf a}^t}^{-1}$. For $PSL(2,Z)_T$ transformations ${\bf a}$ is given
by
\begin{equation}
{\bf a}=\pmatrix{d&0&c&0\cr 0&a&0&b\cr b&0&a&0\cr 0&c&0&d}
\label{newbast}
\end{equation}
while for $PSL(2,Z)_U$, ${\bf a}$ is obtained by interchanging the last
two columns and rows. Finally $T,U$ interchange corresponds to
\begin{equation}
{\bf a}=\pmatrix{{\bf 1}&0\cr 0&\sigma_1\cr}\qquad\qquad
\sigma_1=\pmatrix{0&1\cr 1&0\cr}
\label{newbastu}
\end{equation}
These matrices ${\bf a}$ are $O(2,2;Z)$ matrices which preserve the metric $M$
\begin{equation}
M=\pmatrix{\sigma_1&0\cr 0&-\sigma_1\cr}
\label{metric}
\end{equation}

As explained in the last section,
when one includes the one loop correction to the prepotential $f$, the
$O(2,2;Z)$ group is replaced by
the monodromy group $G$ generated by the three elements
$g_1$, $g_2$ and $\sigma$. The action of these elements on $f$ and $S$
is given by equations (\ref{transform}), (\ref{poly}) and (\ref{stran}).
In the new symplectic basis introduced above, these transformations
act linearly with the upper off-diagonal block ${\bf b}=0$, that is
they are of the form:
\begin{equation}
\pmatrix{{\bf a}&{0}\cr {{\bf a}^t}^{-1}{\tilde{\bf c}}&{{\bf a}^t}^{-1}
\cr}
\label{symp2}
\end{equation}
The matrices ${\bf a}$, ${\tilde{\bf c}}$ for the three generators are
as follows:
\begin{eqnarray}
g_1:~~{\bf a}&=& \pmatrix{0&0&1&0\cr 0&0&0&-1\cr -1&0&0&0\cr 0&1&0&0\cr}\qquad
{\tilde{\bf c}}={\bf 0}\nonumber\\
g_2:~~{\bf a}&=& \pmatrix{1&0&1&0\cr 0&0&0&-1\cr -1&0&0&0\cr 0&1&0&1\cr}\qquad
{\tilde{\bf c}}=\pmatrix{-4&0&0&0\cr 0&0&0&0\cr 0&0&4&0\cr 0&0&0&0\cr}
\label{BAA'}\\
\sigma:~~{\bf a}&=&\pmatrix{1&0&0&0\cr 0&1&0&0\cr 0&0&0&1\cr 0&0&1&0\cr}\qquad
{\tilde{\bf c}}=\pmatrix{0&-1&2&-2\cr -1&0&-2&2\cr 2&-2&4&-1\cr -2&2&-1&0\cr}
\nonumber
\end{eqnarray}

%\begin{equation}
%\matrix{
%F_0&\rightarrow&F_0+2A_{00}X^0+{1\over2}A_{11}X^1+A_{10}X^2+A_{01}X^3\cr
%F_1&\rightarrow&F_1+{1\over 2}A_{11}X^0+2A_{22}X^1+A_{21}X^2+A_{12}X^3\cr
%F_2&\rightarrow&F_2+A_{10}X^0+A_{21}X^1+2A_{20}X^2+{1\over2}A_{11}X^3\cr
%F_3&\rightarrow&F_3+A_{01}X^0+A_{12}X^1+{1\over2}A_{11}X^2+2A_{02}X^3\cr}
%\label{newxf}
%\end{equation}

Note that the matrices $\tilde{\bf c}$ are symmetric
and satisfy Tr$M{\tilde{\bf c}}=0$, where
$M$ is the metric (\ref{metric}).

The abelian group $H$ introduced in (\ref{He}) is generated by symplectic
matrices (\ref{symp2}) with ${\bf a}={\bf 1}$, and $\tilde{\bf c}$:
\begin{equation}
{\tilde{\bf c}}=\sum_i 2N_ig_i^t\pmatrix{\sigma_1&0\cr 0&\sigma_1-2\cr}g_i
\label{newH}
\end{equation}
where $g_i$ can be chosen for instance
as $PSL(2,Z)_T$ matrices of the form (\ref{newbast}).
Since $g_i$ preserve the metric $M$ it is clear that the symmetric matrices
(\ref{newH}) are traceless with respect to $M$. Therefore, by suitable
choices of $g_i$'s and $N_i$'s one can generate all symmetric
$4\times 4$ matrices which are
traceless with respect to $M$, and therefore depending on 9 integer parameters.
They form the 9-dimensional representation of $O(2,2;Z)$ corresponding to
the second rank symmetric traceless tensors, as explained in the last section.

The full perturbative monodromy group contains also the axionic shift $D$
(\ref{sshift}) which in the above symplectic basis corresponds to
\begin{equation}
D:~~\pmatrix{{\bf 1}&0\cr -\lambda M&{\bf 1}\cr}\ ,
\label{newsshift}
\end{equation}
which commutes with the above matrices of $G$, as expected.
The parameter $\lambda$ should also be quantized at the non-perturbative
level. In this way one generates all possible symmetric $4\times 4$ lower
off-diagonal matrices depending on 10 integer parameters, the trace part being
generated by $M$ in (\ref{newsshift}). The full monodromy group is generated
by the 4 generators $g_1$, $g_2$, $\sigma$ and $D$.

\section{Generalization to arbitrary $(4,4)$ compactifications}

The heterotic string compactified on $T^2\times K_3$ with spin connection
identified with the gauge connection gives rise to $N{=}2$ supersymmetry
having,
besides the $U(1)^2$ associated with the dilaton and the graviphoton, a rank 17
gauge group $E_7\times E_8\times U(1)^2$.\footnote{For special points in the
hypermultiplet moduli space, as for example orbifold point of $K_3$, there
could
be extra massless vector multiplets increasing the rank of the gauge group. We
will discuss such situations in the next section.} There are also 20 massless
hypermultiplets in the {\bf 56} representation of $E_7$. In the previous
sections
we discussed the dependence of the prepotential on the $U(1)^2$ vector
multiplets
corresponding to the moduli of the 2-torus $T^2$. However, the complete moduli
space also includes the $2\times 15$ Wilson lines which enlarge the lattice
deformations to $O(2,17)$. At a generic point of this moduli space the gauge
group is broken to $U(1)^{17}$ and all charged hypermultiplets become massive.
Complex co-dimension 1  singularities in the moduli space correspond either to
the appearance of two extra massless vector multiplets which enlarge one of the
$U(1)$ factors to $SU(2)$, or to massless hypermultiplets. These are the
analogues of the $T=U$ singularities discussed in the previous
sections. There are of course higher co-dimensional surfaces analogous
to $T=U=i$ or $\rho$, which correspond to larger gauge groups and/or more
massless hypermultiplets; they are not relevant in the following discussions.

At the classical level, the duality group is $O(2,17;Z)$ which leaves the mass
spectrum and the interactions invariant. This is a subgroup of the symplectic
transformations $Sp(38;Z)$ mentioned in the introduction. As in the last
section, one can choose a field basis in which these
transformations are linear and block diagonal at the tree level. For
convenience we will choose here a basis
\cite{ceres} such that $O(2,17;Z)$ leaves invariant
the diagonal metric
$\eta = {\rm diag}(-1,-1;1,1,\dots,1)$:
\begin{eqnarray}
X^I &=&(X^0,X^1,X^\alpha)\, , ~~~~~X^I X^J \eta_{IJ} =0 \nonumber\\[-6mm]
& & \label{basis}\\[-7mm]
F_I &=& S \eta_{IJ} X^J \nonumber
\end{eqnarray}
where $\alpha=2,\dots,18$ and $S$ is the dilaton. The 17 physical coordinates
$y^{\alpha}$ of the $O(2,17)/$\linebreak $(O(2)\times O(17))$ manifold are
given in terms of
$X$'s by $X^\alpha/X^0 =2y^{\alpha}/(1+y_{\alpha}^2)$. $X^I$ and $F_I$ satisfy
the constraints: $F_I \eta^{IJ} F_J = F_I X^I= 0$. Note that in this basis the
prepotential does not exist, i.e. $F_I$ is not $I$-th derivative of a function.
This is exactly as in the case of $O(2,2)$ in the new basis introduced in
section 4, where the role of $X^s$ and $F_s$ was interchanged to diagonalize
the $O(2,2;Z)$ transformations. If one wishes, one could go back to a basis
where a prepotential exists. The tree-level K\"ahler potential is given by
\begin{equation}
K^{(0)}=-\log i(X^I{\bar F}_I-\xbar^I F_I) = -\log i({\bar S}-S) - \log
X^I\eta_{IJ} {\bar X}^J
\label{kahler2}
\end{equation}
and the $O(2,17)$ transformations in the symplectic basis (\ref{basis}) take
the
form:
\begin{equation}
\pmatrix{X^I\cr F_I}\rightarrow\pmatrix{{\bf a}&0\cr 0&{{\bf a}^t}^{-1}}
\pmatrix{X^I\cr F_I}\ ,
\label{o2n}
\end{equation}
where ${\bf a}$ is a $O(2,17)$ matrix which preserves the metric $\eta$.

The BPS mass formula \cite{ceres} is
\begin{equation}
m = e^{K/2}|n_I^{(e)}X^I - n_{(m)}^I F_I |\ ,
\label{mass}
\end{equation}
which is invariant under K\"ahler transformations. Here $n^{(e)}$ and $n_{(m)}$
are the electric and magnetic charge vectors. The elementary string states have
$n_{(m)}=0$ and $n^{(e)}$ lie in a lattice $\Gamma^{(e)}$ which for instance
can be choosen to be the product of an even self-dual lattice $\Gamma^{(2,2)}$
corresponding to the two-torus with the weight lattices of $E_7\times E_8$.
For convenience we will choose for $\Gamma^{(2,2)}$ the $SO(4)\times
SO(4)$ weight lattice with the conjugacy classes of the two factors being
identified.\footnote{Here we normalise roots to have length $\sqrt 2$.} The
conjugacy class of the scalar in $E_7$ corresponds to the vector multiplets
while the one of {\bf 56} corresponds to hypermultiplets. In fact, for
$n_{(m)}=0$, the mass (\ref{mass}) is just the left moving momentum of the
two-torus $|p_L|$, i.e. they correspond to the ground state of left-moving
sector with momentum $p_L$. Massless states are the ones with $m=0$ and
$n^{(e)}_I\eta^{IJ} n^{(e)}_J = 2$ for vector multiplets and =$3/2$ for
hypermultiplets. Thus the point $y^{\alpha}=0$ corresponds to the gauge group
$E_7\times E_8 \times SO(4)$ with massless hypermultiplets in {\bf 56}
representation of $E_7$ whose multiplicity is governed by the cohomology of
$K_3$ and is $20$. On the other hand it is clear from the constraints for
massless states that at generic values of $y^{\alpha}$'s, there are no charged
massless states and therefore the gauge group is $U(1)^{17}$. The symmetry
group $O(2,17;Z)$ is the automorphism group of $\Gamma^{(e)}$.

The complex co-dimension 1 surface of singularity corresponding to the
enhancement
of one of the $U(1)$'s to $SU(2)$ (i.e. when two charged vector multiplets
become massless) is defined by the equation
\begin{equation}
n_I^{(e)}X^I=0
\label{sing}
\end{equation}
for a particular choice of $n^{(e)}$ vector obeying $n^{(e)}_I\eta^{IJ}
n^{(e)}_J = 2$. Different choices of such charge vectors define different
surfaces of singularity and they are related to different $U(1)$'s being
enhanced to $SU(2)$. For different vectors $n^{(e)}$'s that are related by
$O(2,17;Z)$ transformation, the corresponding surfaces are also $O(2,17;Z)$
transforms of each other. Similarly, the singular surfaces associated with the
appearance of massless hypermultiplets are given by eq.(\ref{sing}) with
$n^{(e)}_I\eta^{IJ} n^{(e)}_J = 3/2$. The appearance of these massless states
gives rise to logarithmic singularities in the prepotential as in the $O(2,2)$
case discussed previously. In the following we will identify the coefficient of
these logarithmic singularities as they enter in the monodromy matrices.

Let us denote by $f_I$ the one-loop corrections to $F_I$ of
eq.(\ref{basis}). The one-loop correction to the K\"ahler potential is
\begin{equation}
K^{(1)}=-{1\over S-\sbar}{(X^I{\bar f}_I-\xbar^If_I)\over
X^K\eta_{KL}\xbar^L}\ .
\label{Kahler1}
\end{equation}
Consider now the behaviour of $K^{(1)}$ near a singular surface
$n_I^{(e)}X^I=0$. The direction orthogonal to the surface, and subject to the
constraint $X^I X^J \eta_{IJ} =0$, is $\delta X^I=\eta^{IJ}n_J^{(e)}\epsilon$,
where $\epsilon$ is an infisitesimal parameter. We are interested in the
component of the metric along this direction, since it is this component which
has a logarithmic singularity near the surface. Expanding the K\"ahler
potential
(\ref{kahler2}) and  (\ref{Kahler1}) in powers of $\epsilon$ and $\bar\epsilon$
and extracting the coefficient of $\epsilon\bar\epsilon$, one finds:
\begin{eqnarray}
G^{(1)}_{\epsilon\bar\epsilon} &=& {i\over 2} G^{(0)}_{\epsilon\bar\epsilon}
[{1\over{n^{(e)}}^2}n^{(e)}_I\eta^{IJ}(\delta_\epsilon
f_J-\delta_{\bar\epsilon}
{\bar f}_J) +{X^I{\bar f}_I-\xbar^If_I\over X^K\eta_{KL}\xbar^L}]\nonumber\\
G^{(0)}_{\epsilon\bar\epsilon} &=& -{{n^{(e)}}^2\over X^I\eta_{IJ}\xbar^J}
\label{Kahlerexp}
\end{eqnarray}
where ${n^{(e)}}^2\equiv n^{(e)}_I\eta^{IJ}n^{(e)}_J$. Note that the tree-level
metric $G^{(0)}$ does not mix the $\epsilon$ direction with the directions
tangential to the singular surface since the linear terms in the expansion of
$K^{(0)}$ vanish on the surface. The linear terms in the expansion of $K^{(1)}$
are proportional to
\begin{equation}
{\epsilon\over X^K\eta_{KL}\xbar^L}
[n^{(e)}_I\eta^{IJ}{\bar f}_J-\xbar^I\delta_\epsilon f_I - c.c.]
\label{linear}
\end{equation}
We know that the one-loop metric near the singular surface has a logarithmic
singularity of the form
$G^{(1)}_{\epsilon\bar\epsilon}/G^{(0)}_{\epsilon\bar\epsilon}={c\over \pi}
\log |{n^{(e)}_IX^I\over X^0}|^2$ with $c={n^{(e)}}^2=2$ for vector multiplets,
and $c=-10{n^{(e)}}^2=-15$ for hypermultiplets. The appearance of ${n^{(e)}}^2$
can be understood from the fact that these are the square of the charges of the
states that become massless with respect to the $U(1)$ defined by the
$\epsilon$-direction. The particular values 2 and 15 are associated with
charges $\pm 1$ for the $SU(2)$ adjoint representation, and
$\pm{\sqrt 3}/2$ for the 20 hypermultiplets. As mentioned before, the
factor 10 is related to the cohomology of $K_3$, and $O(2,17)$
deformations do not alter this value. As for the mixed components of the
one-loop metric involving $\epsilon$ and a direction tangential to the surface,
there is no logarithmic singularity since the sum over the charges vanishes.
These requirements together with eqs.(\ref{Kahlerexp}) and (\ref{linear}) imply
that the singular part of $f_I$ near the surface is:
\begin{equation}
f_I=-{2iN\over \pi}n^{(e)}_In^{(e)}_J X^J\log {n^{(e)}_LX^L\over X^0}
\label{fI}
\end{equation}
where $N=1$ or $-10$ for the case of vector multiplets or hypermultiplets,
respectively.

The presence of logarithms in $f_I$ modifies the classical monodromies just
as in
the $O(2,2;Z)$ case. The analogue of the $T\leftrightarrow U$ exchange
corresponds
now to the Weyl reflections $W_{n^{(e)}}$
defined by the vectors $n^{(e)}$'s satisfying
${n^{(e)}}^2=2$ (i.e. for the vector multiplets).
$W_{n^{(e)}}$ is an automorphism of
the charge lattice and, at the classical level,
it satisfies $(W_{n^{(e)}})^2=1$. However at the
quantum level this relation is no longer true
due to the logarithmic singularities in $f$, as
in the $O(2,2)$ case. Indeed, $(W_{n^{(e)}})^2\equiv Z_{n^{(e)}}$
corresponds to moving a point around the singular
surface $n^{(e)}_IX^I=0$.
Consider a vector $n^{(e)}$
lying in the $\alpha$-directions. From equation (\ref{fI})
it is easy to see that $(W_{n^{(e)}})^2$
shifts $F_I$ as $F_I\rightarrow F_I+ 4
n^{(e)}_In^{(e)}_J X^J$. This results in the following symplectic
transformation:
\begin{equation}
Z_{n^{(e)}}=\pmatrix{{\bf 1}&0\cr{{\tilde{\bf c}}^v}
&{\bf 1}}  \qquad \qquad {\tilde{\bf c}}^v=4n^{(e)}{n^{(e)}}^t
\label{zne}
\end{equation}
It follows that $W_{n^{(e)}}$ must be of the form:
\begin{equation}
W_{n^{(e)}}=\pmatrix{{\bf a}&0\cr {{\bf a}^t}^{-1}{\tilde{\bf c}}&{{\bf
a}^t}^{-1}}
\label{weylr}
\end{equation}
where ${\bf a}$ is the element of $O(2,17;Z)$ corresponding to the above Weyl
reflection and ${\tilde{\bf c}}$ is a symmetric matrix satisfying
the condition ${{\bf a}^t}^{-1}{\tilde{\bf c}}{\bf a}+\tilde{\bf c}=
-4n^{(e)}{n^{(e)}}^t$.

In the case of ${n^{(e)}}^2=3/2$ corresponding to ${\bf 56}$ of $E_7$ (i.e. for
hypermultiplets) the reflection is not a symmetry of the lattice. However there
is still a non-trivial monodromy
$Z_{n^{(e)}}$ associated with moving a point around such singular surfaces:
\begin{equation}
Z_{n^{(e)}}=\pmatrix{{\bf 1}&0\cr{{\tilde{\bf c}}^h}
&{\bf 1}}, \qquad\qquad {\tilde{\bf c}}^h=40n^{(e)}{n^{(e)}}^t
\label{zneh}
\end{equation}
where the coefficient 40 appears due to the multiplicity
20 of the hypermultiplets that become massless.

Similarly to the $O(2,2)$ case discussed in sections 3 and 4, the fact
that $(W_{n^{(e)}})^2$ is not equal to the identity implies that the classical
group $O(2,17;Z)$ is replaced by a quantum monodromy group $G$. The latter
is defined by the fundamental group of the space
obtained after removing the singular
surfaces from the fundamental domain of $O(2,17;Z)$
in $O(2,17)/O(2)\times O(17)$.
Note that the number of singular surfaces in the fundamental domain
is given by the number of distinct $O(2,n;Z)$ orbits
among the lattice vectors satisfying $(n^{(e)})^2=2$ or $3/2$ and is finite.
The fundamental group is finitely presented, and when $Z_{n^{(e)}}$ are set
equal to identity, this group reduces to $O(2,17;Z)$. The subgroup generated by
$Z_{n^{(e)}}$'s defines a normal abelian subgroup $H$ of $G$. In the symplectic
basis an arbitrary element of $H$ is given by
\begin{equation}
\pmatrix{{\bf 1}&0\cr{\tilde{\bf c}}&{\bf 1}}\qquad\qquad
{\tilde{\bf c}}=\sum_iN_i g_i^t {\tilde{\bf c}}^v g_i
+\sum_jM_j g_j^t {\tilde{\bf c}}^h g_j
\label{Hgroup}
\end{equation}
where $g_i$ are $O(2,17;Z)$ elements.
In this way, we generate a general
symmetric matrix ${\tilde{\bf c}}$ depending on $19\times 20/2$ integer
parameters. It is decomposed into a sum of two irreducible representations of
$O(2,17)$: a traceless symmetric tensor and a singlet corresponding to the
trace.
Note that the latter can be identified with the quantized dilaton shift having
the form:
\begin{equation}
\pmatrix{{\bf 1}&0\cr{\eta}&{\bf 1}}
\label{sshiftn}
\end{equation}
Of course at the perturbative level, on top of this transformation one can add
an arbitrary dilaton shift with $\eta$ replaced by $\lambda\eta$.
The quotient group $G/H$ is isomorphic to $O(2,17;Z)$. A representative element
in a class of $G/H$ is given in the symplectic basis as:
\begin{equation}
\pmatrix{{\bf a}&0\cr {{\bf a}^t}^{-1}{\tilde{\bf c}}&{{\bf
a}^t}^{-1}}
\label{goverh}
\end{equation}
where ${\bf a}$ is the corresponding $O(2,17;Z)$ matrix and ${\tilde{\bf c}}$
is some symmetric matrix, whose precise form is determined by the relations
satisfied by the generators of $G$ as was done in the case of $O(2,2)$ in
sections 3 and 4. For example, as stated above for the Weyl reflections
$W_{n^{(e)}}$, ${\tilde{\bf c}}$ is constrained
by the group relation $(W_{n^{(e)}})^2=Z_{n^{(e)}}$.
Unfortunately at present we do not know the complete set of group
relations defining the fundamental group and therefore we are unable
to construct the ${\tilde{\bf c}}$'s for various generators explicitly.
For consistency at the non-perturbative level the entries of ${\tilde{\bf c}}$
must be quantized such that ${\tilde{\bf c}}\Gamma_{(m)}\subset\Gamma^{(e)}$,
where $\Gamma_{(m)}$ is the magnetic charge lattice
which, as we shall discuss in the next
section, is the lattice dual to $\Gamma^{(e)}$
with respect to the metric $\eta$.
One can see that the ${\tilde{\bf c}}$'s appearing in $H$ subgroup
(\ref{Hgroup}) satisfy this condition. Although we are unable to determine
$G$ completely, we can however say that it is some finite index subgroup of the
group of matrices of the form (\ref{goverh}) with ${\bf a}\in O(2,17;Z)$ and
${\tilde{\bf c}}$ an arbitrary symmetric matrix satisfying the quantization
condition.

\section{$(4,0)$ models}

So far we have discussed generic $(4,4)$ models leading to rank $r=17$ gauge
group. However in the moduli space of hypermultiplets, there are special points
where additional vector multiplets become massless leading to an increase in
the
rank. For example at the $Z_2$ orbifold point one gets an extra $SU(2)$ factor
increasing the rank to 18, while for special radii one can even get rank 22
gauge groups. At these special points the moduli space of vectors is usually
increased to $O(2,r)/(O(2)\times O(r))$ and the classical symmetry group is
$O(2,r;Z)$. The above analysis can again be repeated. We first introduce the
symplectic basis $(X^I, F_I)$ with $I=0,1,\dots r+1$ and $X^I\eta_{IJ}X^J=0$ on
which the $O(2,r;Z)$ transformations act linearly by block diagonal symplectic
matrices. The mass spectum is again given as in eq.(\ref{mass}) with the charge
vectors $n^{(e)}$ living in a lattice $\Gamma^{(2,r)}$. We assume for
simplicity
that the sublattice $\Gamma_v$ associated with the charges of vector
multiplets is
even and integral, which is the case for orbifolds. For orbifolds, it is also
true that the full lattice $\Gamma^{(2,r)}$ is the dual of $\Gamma_v$, the
non-trivial conjugacy classes $C$ of $\Gamma^{(2,r)}$ with respect to
$\Gamma_v$
being associated with hypermultiplets. In the full string theory, each of these
classes is coupled to a block of the internal conformal field theory which
describes the remaining $(22-r)$ right movers. The data from the latter which
is relevant here, is the multiplicity $m_C$ of the number of operators in
the Neveu-Schwarz sector carrying conformal dimension $(1/2,\Delta_C)$ with
$\Delta_C\le 1$ in the block coupled to the conjugacy class $C$. Of course,
world-sheet modular invariance implies that $\Delta_C +{1\over 2}{n^{(e)}}^2$
is an integer for $n^{(e)}$ belonging to the class $C$. Obviously $m_C$ and
$\Delta_C$ do not change under $O(2,r)$ deformations. This is similar to the
multiplicity 20 of the $\bf 56$'s of $E_7$ in the $(4,4)$ models. The
classical symmetry group which should preserve the spectrum is $O(2,r,Z)$ which
preserves the lattice $\Gamma_v$. At a generic point in the moduli space
$O(2,r)/(O(2)\times O(r))$,
the gauge group is $U(1)^r$ and there are no massless hypermultiplets.

At the one loop level the prepotential again develops logarithmic
singularities near complex co-dimension 1 surfaces where
extra massless particles
appear. The ones associated with the enhancement of gauge symmetry to
$U(1)^{r-1}
\times SU(2)$ are given by the surfaces $n^{(e)}\cdot X=0$ for $n^{(e)} \in
\Gamma_v$ and ${n^{(e)}}^2=2$; the ones associated with the appearance of extra
massless hypermultiplets correspond
to $n^{(e)}\cdot X=0$ with $n^{(e)}$  belonging to a
non-trivial class $C$ in $\Gamma^{(2,r)}$ with ${n^{(e)}}^2+2\Delta_C =2$. As
in
the (4,4) case, one can show that the singular part of $f_I$'s near such a
surface is given by eq.(\ref{fI}), with $N$ being 1 for vector multiplets and
$-m_C$ for hypermultiplets associated with the conjugacy class $C$. As before
the presence of logarithmic singularity gives rise to non-trivial monodromies.
The Weyl reflections associated with $n^{(e)} \in \Gamma_v$ satisfying
${n^{(e)}}^2=2$ are again represented by the matrices $W_{n^{(e)}}$ of
eq.(\ref{weylr}). Similarly $W_{n^{(e)}}^2 \equiv Z_{n^{(e)}}$ is given
by (\ref{zne}). For hypermultiplets the reflections are not automorphisms of
the lattice. However moving a point around such surfaces one gets monodromies
that are given by the matrices $Z_{n^{(e)}}$ of eq.(\ref{zneh})
with ${\tilde{\bf
c}}^h= 4 m_C n^{(e)}{n^{(e)}}^t$ for $n^{(e)}$ in the conjugacy class $C$.
The normal abelian subgroup $H$ consists of
elements of the type (\ref{Hgroup}) with
$\tilde{\bf c}=\sum_i N_i g_i^t {\tilde{\bf c}}^v g_i
+\sum_j M_j g_j^t {\tilde{\bf c}}^h g_j$,
where $g_i$ are $O(2,r;Z)$ elements.
In this way, we generate a general
symmetric matrix ${\tilde{\bf c}}$ depending on $(r+2)(r+3)/2$ integer
parameters. It is decomposed into a sum of two irreducible representations of
$O(2,r)$: a traceless symmetric tensor and a singlet corresponding to the
trace. The latter is identified with quantized axionic shift as before.
The quotient $G/H$ is isomorphic to
$O(2,r,Z)$ and a general element of $G$ is again of the form given in
eq.(\ref{goverh}) where ${\tilde{\bf c}}$ is to be determined from the
precise form of the relations defining the fundamental group.

Now let us discuss the consistency of the monodromy group when non-perturbative
effects are taken into account. This means that the monodromy preserves the
complete mass spectrum of BPS states involving electric as well as magnetic
charges. The monodromy group $G$ acts as symplectic transformation of electric
and magnetic charge vectors $(n^{(e)}, n_{(m)})$. Dirac quantization condition
for magnetic charges implies that magnetic charge vectors must be in the
dual lattice of electric charge vectors $\Gamma^{(2,r)}$. This means that
magnetic charges in fact lie in $\Gamma_v$. A general element of the
perturbative monodromy group we have discussed so far consists of matrices
whose upper off-diagonal block is zero. Morever the diagonal blocks are made
up of $O(2,r,Z)$ matrices which by definition preserve $\Gamma_v$ and
therefore the electric and magnetic charge lattices separately. The
non-trivial question is whether the lower off-diagonal block $\tilde{\bf c}$
which mixes the magnetic charge lattice with the electric charge one, is
consistent. In other words we must have $\tilde{\bf c} n_{(m)} \in
\Gamma^{(2,r)}$. Since $\tilde{\bf c}$ appearing in $H$
is made up of matrices of
the form $2 n^{(e)} {n^{(e)}}^t$ this condition is obviously satisfied.
${\tilde{\bf c}}$ appearing in a general element of $G$ must also
satisfy this condition. Thus we see that again $G$ is a finite index
subgroup of the group of matrices of the form (\ref{goverh}) with
${\bf a}\in O(2,r;Z)$ and ${\tilde{\bf c}}$ an arbitrary symmetric matrix
satisfying the quantization condition.
The non-perturbative
consistency also implies the quantization of
the dilaton shift: $S\rightarrow S\,$+
integer.

To illustrate the above let us consider $Z_2$ orbifold and restrict to a
subspace of two moduli which generalize the $O(2,2)$ case discussed in sections
2, 3 and 4. More precisely, we start with a model defined from the usual
toroidal compactification $T^2\times T^4$ of the $E_8\times E_8$ heterotic
theory by a $Z_2$ twist on the $T^4$ together with a $Z_2$ shift $\delta$
acting on the $\Gamma^{(2,2)}$ momentum lattice corresponding to
$T^2$. In order to satisfy the level matching condition $\delta^2$ must be 1/2.
Note that this is in contrast with the usual orbifold constructions where the
shift is embedded in one of the $E_8$ factors breaking it to $E_7\times SU(2)$.
Now the gauge group is $E_8\times E_8\times U(1)^2$ at a generic point in the
moduli space of $T^2$. In terms of the integers $n_i, m_i$ that define the
momenta (\ref{pL}), (\ref{pR}), the effect of this shift is the following. In
the untwisted sector, vector multiplets are associated with $m_2+n_1$ even
integers, while hypermultiplets correspond to $m_2+n_1$ odd. In the twisted
sector $m_2$ and $n_1$ are shifted by 1/2 and these states are hypermultiplets.
With respect to the lattice $\Gamma_v$ corresponding to $m_2+n_1$ even, the
charge lattice has now four classes. Besides the trivial class $C_0$,
the non-trivial ones are $C_1$ associated
with $m_2, n_1\in Z$ and $m_2+n_1$ odd, and $C_2$ and $C_3$
associated with $m_2,
n_1\in Z+1/2$ and $m_2+n_1$ even and odd, respectively. The data from the
remaining conformal field theory $(m_C,\Delta_C)$ discussed above is (1,0)
for $C_1$, (32,3/4) for $C_2$ and (8,1/4) for $C_3$. Furthermore the
tree-level symmetry group ${\tilde O}(2,2;Z)$ is a subgroup of $O(2,2;Z)$
defined
in section 4, which leaves these classes invariant. More precisely, its even
part is the subgroup of $SL(2,Z)_T\times SL(2,Z)_U$ obtained by identifying the
cosets of the two factors with respect to the $\Gamma(2)$ subgroup of
$SL(2,Z)$;
its odd part is obtained by including  the $T\leftrightarrow U$ exchange.

Repeating the analysis of section 2, one can show that the third derivative of
the one-loop prepotential $f_{TTT}$ is given as a sum of contributions from the
four classes, each of them being expressed by the r.h.s.\ of eq.(\ref{fTTT}):
\begin{equation}
f_{TTT}=4\pi^2 \frac{U-\ubar}{(T-\tbar)^2} \sum_{C_{\ell}}
\int{d^2\tau}\,\bar{F}_{\ell}(\bar{\tau})\,
\sum_{p_L,p_R\in C_{\ell}}p_L\bar{p}_{R}^3\, e^{\pi i\tau\, |p_{L}|^2}\,
e^{-\pi i \bar{\tau}\, |p_{R}|^2}\, .
\label{fTTT2}
\end{equation}
At $\tau_2\rightarrow\infty$, $2i\pi^2\bar{F}_{\ell}$ behaves as ${\bar
q}^{-1}$
for $\ell =0$, $-1{\bar q}^{-1}$ for ${\ell}=1$, $-32{\bar q}^{-1/4}$ for
${\ell}=2$ and $-8{\bar q}^{-3/4}$ for ${\ell}=3$, where $q=e^{2i\pi\tau}$. One
can verify from eq.(\ref{fTTT2}) that in each class there is a
simple pole
singularity associated with the appearance of massless states. The condition
$p_L=0$ gives the lines $m_1+m_2U+n_1T+n_2TU=0$ while the massless condition
for the right movers gives $m_1n_2-m_2n_1=1,1,1/4,3/4$ for the four classes
$C_0,C_1,C_2,C_3$, respectively. For $C_0$ there are four distinct singular
lines (modulo the automorphism group) $T=U$, $T=U+1$, $T=-1/U$
and $T=U/(U+1)$, where the
gauge group becomes $SU(2)\times U(1)$. For the other classes there is one
representative singular line each which we can choose to be $T=-1/(U+1)$ for
$C_1$, $T=U$ for $C_2$ and $T=3U$ for $C_3$, where we have two massless
hypermultiplets. Note that the singular line of class $C_2$ coincides with one
of the lines for $C_0$ implying that the two massless hypermultiplets come in
one $SU(2)$ doublet.

To each of the above singular lines there is an associated non-trivial
monodromy.
For the $T=U$ singularity, where besides
the $SU(2)$ gauge symmetry also 32 massless $SU(2)$ doublet hypermultiplets
appear, we have the following monodromy for $f$:
\begin{eqnarray}
T &{\rm around}& U:\qquad\qquad\qquad f\rightarrow f-62(T-U)^2,
\label{ftran3}
\end{eqnarray}
where the coefficient $-62$ is due to the contribution +2 of
the vectors and $-64$ of the hypermultiplets.
For the other 3 $SU(2)$ lines the monodromies are:
\begin{eqnarray}
T&{\rm around}& (U+1):\qquad\qquad f\rightarrow f+2(T-U-1)^2 \nonumber\\
T&{\rm around}& -{1\over U}:
\qquad\qquad\qquad f\rightarrow f+2(1+TU)^2\nonumber\\
T&{\rm around}& {U\over U+1}:\qquad\qquad f\rightarrow f+2(TU+T-U)^2.
\label{ftran2}
\end{eqnarray}
Finally, for the remaining two hypermultiplet lines we have:
\begin{eqnarray}
T&{\rm around}& 3U:\qquad\qquad f\rightarrow f-16(T-3U)^2 \nonumber\\
T&{\rm around}& -{1\over U+1}:
\qquad\qquad\qquad f\rightarrow f-2(TU+T+1)^2.
\label{ftran4}
\end{eqnarray}
Here, we have used the particular values for the multiplicities of the
various classes to get the multiplicative coefficients.

The general element of the monodromy group $G$ will be a representation
of the fundamental group of the underlying space, however the latter cannot
be realized as a braid group due to the presence of $T=3U$ singular line.
For this reason, we consider a slight modification of the above example where
we
define the $Z_2$ orbifold group by including a simultaneous $Z_2$ shift in
$E_8 \times E_8$ lattice in such a way that the level matching condition
is satisfied. For instance we can shift by a vector of $O(16)$ in one of the
$E_8$'s. The shift in the $T^2$ torus part is the same as before. As a
result in the twisted sector the remaining conformal field theory provides
a minimum right moving dimension $={1\over 4} + {1\over 2}={3\over 4}$ where
the ${1\over 2}$ appears due to the extra shift in $E_8$. This means that
in class $C_3$ now there is no massless state and hence the $T=3U$ singularity
disappears.

The conformal data $(\Delta_C, m_C)$ = $(0,1)$ for the class $C_0$,
$(0,1)$ for $C_1$ class, $({3\over4},m)$ for $C_2$ class and $({5\over 4},n)$
for $C_3$ class. Here $m$ and $n$ are some integers that depend on the
precise form of the shift vector in $E_8\times E_8$ (e.g. when the shift is a
vector of $O(16)$, $m=128$). The singular lines for $C_0$ are as before
(modulo the automorphism group) $T=U$, $T=U+1$, $T=-1/U$
and $T=U/(U+1)$, where the
gauge group becomes $SU(2)\times U(1)$.
Similarly for classes $C_1$ and $C_2$ the singular lines $T=-1/(U+1)$
and $T=U$, respectively where two massless charged hypermultiplets appear,
remain unchanged. The difference now is that for class
$C_3$ there is no singular line. As before, the singular line of class $C_2$
coincides with one of the lines for $C_0$ implying that the two massless
hypermultiplets come in one $SU(2)$ doublet. Monodromies of the prepotential
$f$
around these singularities are:
\begin{eqnarray}
T&{\rm around}& U:\qquad\qquad\qquad ~~f\rightarrow f-2(m-1)(T-U)^2 \nonumber\\
T&{\rm around}& (U+1):~\qquad\qquad f\rightarrow f+2(T-U-1)^2 \nonumber\\
T&{\rm around}& -{1\over U}:
\qquad\qquad\qquad f\rightarrow f+2(1+TU)^2\nonumber\\
T&{\rm around}& {U\over U+1}:\qquad\qquad\quad f\rightarrow f+2(TU+T-U)^2.
\nonumber\\
T&{\rm around}& -{1\over U+1}:
{}~\qquad\qquad f\rightarrow f-2(TU+T+1)^2.
\label{ftrann1}
\end{eqnarray}
These transformations together with all conjugations generate the normal
abelian
subgroup $H$.

Our aim now is to determine the action of the full monodromy
group $G$. To this end we note that although the above singular lines are
inequivalent with respect to the automorphism group ${\tilde O}(2,2;Z)$, they
are mapped to each other by the action of $O(2,2;Z)$. The prepotential $f$,
however, is mapped to different prepotentials under this action as the right
hand side of the differential equation (\ref{fTTT2}) transforms non-trivially.
Let us define $f_1 \equiv f$, $f_2 (T,U) = T^2 f(-{1\over T},U)$,
$f_3(T,U)= (T+1)^2 f(-{1\over{T+1}},U)$, $f_4(T,U)= (T-1)^2
f(-{T\over{T-1}},U)$,
$f_5(T,U)= T^2 f(-{{T+1}\over T},U)$ and $f_6(T,U)= f(T-1,U)$. Then $O(2,2;Z)$
acts on ${\bf f}$ as a $6$-dimensional representation, where ${\bf f}$ is a
$6$-column vector whose entries are $f_i$'s. This is of course modulo the
quadratic polynomials in $T$ and $U$ which will give the quantum modification
of the monodromy group $G$. Our strategy now is first to determine the action
of the monodromy group $G'$ on $\bf f$ and then restrict to the subgroup
that does not mix $f_1$ with the other $f_i$'s. The latter therefore determines
the monodromy group $G$.

$G'$ is a representation of the braid group introduced in section 3, with
generators $g_1$, $g_2$ and $\sigma$ satisfying the relations
(\ref{qrelations}). The difference now is that we are looking for a
6-dimensional representation of the braid group. More explicitly, the action of
the generators on $\bf f$ is:
\begin{equation}
g_1 : {\bf f} \rightarrow {1\over T^2}(M_1{\bf f}+ A(T,U)), ~~ g_2 : {\bf f}
\rightarrow {1\over (T+1)^2}(M_2{\bf f}+B(T,U)), ~~ \sigma :
{\bf f} \rightarrow M_{\sigma}({\bf f}+K(T,U))
\label{braidn}
\end{equation}
where $A$,$B$ and $K$ are 6-column vectors with entries being quadratic
polynomials in $T,U$, and the matrices $M_1$, $M_2$ and $M_{\sigma}$ provide a
6-dimensional representation of $O(2,2;Z)$ as follows:
\begin{equation}
M_1 =\pmatrix{0&1&0&0&0&0 \cr 1&0&0&0&0&0 \cr 0&0&0&1&0&0 \cr 0&0&1&0&0&0
\cr 0&0&0&0&0&1 \cr 0&0&0&0&1&0 } ~~ M_2 =
\pmatrix{0&0&0&0&1&0 \cr 0&0&0&1&0&0 \cr 1&0&0&0&0&0 \cr 0&0&0&0&0&1
\cr 0&0&1&0&0&0 \cr 0&1&0&0&0&0 } ~~
M_{\sigma} = \pmatrix{1&0&0&0&0&0 \cr 0&1&0&0&0&0 \cr 0&0&0&0&1&0 \cr
0&0&0&1&0&0 \cr 0&0&1&0&0&0 \cr 0&0&0&0&0&1 }
\label{Mis}
\end{equation}
Finally $\sigma^2 \equiv Z_1$ corresponds to moving a point around the $T=U$
singularity and is given by
\begin{equation}
Z_1: {\bf f}\rightarrow {\bf f} + 2(T-U)^2 P,
\label{Z1new}
\end{equation}
where $P$ is a 6-column vector with constant entries satisfying $P_2=P_4=P_6$
and $P_3=P_5$. In our case these values are $P_1 = (1-m)$, $P_2 = 1$ and
$P_3=-1$. Monodromies around other singular lines are obtained by conjugations.
Note that under conjugations by elements of $O(2,2;Z)$ that are not
in ${\tilde O}(2,2;Z)$ the vector $P$ changes by an appropriate permutation
of its entries. Thus the additive term in $f_1$ will change in accordanace
with the equation (\ref{ftrann1}) as it should.

In order to determine the polynomials $A$, $B$ and $K$ appearing in the
transformations (\ref{braidn}), we first use the freedom to redefine $f_i$'s
by adding quadratic polynomials in $T$ and $U$. One can show that by using this
freedom one can set for example $A=0$ and $B_2 = B_4 = B_6 = 0$ and furthermore
$B_1 = (C_1 T^2 - C_2)$, $B_3 = (C_2 T^2 - C_3)$ and $B_5 = (C_3 T^2 - C_1)$,
where $C_i$'s are quadratic polynomials in $U$. The vector $K$ and $C_i$'s are
then determined by imposing the relations (\ref{qrelations}) of the braid
group. The algebra is cumbersome, but using Mathematica one can show that
the braid relations completely determine the polynomials entering in $K$ and
$C_i$'s in terms of the constants $P_1$, $P_2$ and $P_3$ and are given as:
\begin{eqnarray}
K_1 &=& P_1 (T-U)(2T-TU) + P_2 (T-U)(T+U-TU) + P_3 (T-U)\nonumber\\
K_2 &=& P_1 (T-U) +P_2 (T-U)(T-U-TU+2) -P_3(T-U)TU\nonumber\\
K_3 &=& (P_1-P_3)T +P_2(T-U)(T+U-TU) +P_3 (T-U)(2T-TU+1)\nonumber\\
K_4 &=& -P_1 (T-U)TU + P_2 (T-U)(T-U-TU+2) +P_3(T-U)\nonumber\\
K_5 &=& -(P_1-P_3)U +P_2(T-U)(T+U-TU) +P_3 (T-U)(2T-TU+1)\nonumber\\
K_6 &=& P_2 (T-U)(T-U-TU+2) +P_3(T-U)(1-TU)\nonumber\\
C_1 &=& C_3 = P_2+P_3 \qquad\qquad\qquad C_2 = P_1+P_2
\label{solutionn}
\end{eqnarray}
Finally the monodromy group $G$ is the subgroup of the above representation
of the braid group generated by $g_1$, $g_2$ and $\sigma$, which does not
mix $f_1$ with the other $f_i$'s. More specifically the normal abelian
subgroup $H$ is generated by $Z_1$ of
(\ref{Z1new}) and all conjugations of $Z_1$ by the braid
group elements. The coset $G/H$ being isomorphic to ${\tilde O}(2,2;Z)$ can
be generated by $g_1 \sigma^{-1} g_1 \sigma$, $g_2 \sigma^{-1} g_2 \sigma$,
$(g_2 g_1)^2$ and $\sigma$.

\section{Concluding remarks}

In this paper we studied the perturbative monodromies of the prepotential in
$N{=}2$ heterotic string models
in four dimensions. At the tree-level the duality group is a direct
product of $Z$ corresponding to the dilaton shift with $O(2,r;Z)$ given by the
automorphisms of the charge lattice, where $r$ is the rank of the gauge group.
In some symplectic basis, the duality group acts in a block
diagonal form. At the one-loop level, due to the presence of singularities
associated with the appearance of massless states at complex co-dimension 1
surfaces in the moduli space of vector multiplets, its fundamental group gets
modified. The resulting quantum monodromies associated with closed curves
around
the singular surfaces which acted as identity at the tree-level, now get
modified by a lower off diagonal symmetric matrix which depends on
$(r+2)(r+3)/2$ integer parameters. They define a normal abelian subgroup $H$
of the monodromy group $G$. The quotient group $G/H$ is isomorphic
to the duality group $O(2,r;Z)$.

In order to find the quantum duality group
$G$, it is necessary to find the fundamental group
of the quantum moduli space. We have solved completely
this problem in the $r=2$ case in the $(4,4)$ compactification,
where the fundamental group is known
to be related to the braid group, but for $r\geq 3$
we do not have a complete solution. In section 6, we also gave an example
of $r=2$ case where there are also singularities due to the appearance of
massless hypermultiplets. The monodromy group turns out to be a
subgroup of the braid group where the elements of the latter are realized
in a non-trivial representation.

In view of the recent work of Seiberg and Witten
in the rigid theory, one can ask the question whether at
the non-perturbative level the monodromy group is further modified.
On general grounds we know that a non-perturbative generator
will be an element of $Sp(2r+4, Z)$, with a non-vanishing $\bf b$ entry
(see eq.(\ref{symp})).\footnote{In the $N{=}4$ theory such a generator
is the $Z_2~(S\rightarrow -1/S)$ generator of $PSL(2,Z)_S$.}
The relation of monodromies to braid groups
may be helpful in identifying the non-perturbative monodromy group and in
studying the dynamics of $N{=}2$ superstrings.
\vskip 5mm

{\bf Acknowledgements}

I.A. and T.R.T.\ acknowledge the hospitality of CERN Theory Division.
E.G. and K.S.N. acknowledge the hospitality of the Centre de Physique
Th\'eo\-rique at Ecole Polytechnique, and thank B. Dubrovin and especially
M.S. Narasimhan for discussions. Finally, we thank G. Leontaris for helping us
in the Mathematica program.

\end{document}